\documentclass[twocolumn,superscriptaddress,amsmath,amssymb,aps,sort&compress,prx]{revtex4}

\usepackage[utf8]{inputenc}

\usepackage{times,color,amsthm,graphics,graphicx,bm,bbm,dcolumn}
\usepackage{epsfig}
\usepackage{graphicx}
\usepackage{xcolor}
\usepackage[colorlinks,urlcolor=black,citecolor=blue,linkcolor=black]{hyperref}

\setcitestyle{square}

\begin{document}

\title{Decay of persistent currents in annular atomic superfluids}

\author{K.~Xhani}
\thanks{These authors contributed equally to this work.}
\affiliation{Istituto Nazionale di Ottica del Consiglio Nazionale delle Ricerche (CNR-INO) c/o LENS, 50019 Sesto Fiorentino, Italy}
\email[E-mail to: ] {xhani@lens.unifi.it}

\author{G.~Del Pace}
\thanks{These authors contributed equally to this work.}
\affiliation{Istituto Nazionale di Ottica del Consiglio Nazionale delle Ricerche (CNR-INO) c/o LENS, 50019 Sesto Fiorentino, Italy}
\affiliation{European Laboratory for Nonlinear Spectroscopy (LENS), University of Florence, 50019 Sesto Fiorentino, Italy}
\affiliation{Department of Physics, University of Florence, 50019 Sesto Fiorentino, Italy}

\author{F.~Scazza}
\affiliation{Department of Physics, University of Trieste, 34127 Trieste, Italy}
\affiliation{Istituto Nazionale di Ottica del Consiglio Nazionale delle Ricerche (CNR-INO) c/o LENS, 50019 Sesto Fiorentino, Italy}
\affiliation{European Laboratory for Nonlinear Spectroscopy (LENS), University of Florence, 50019 Sesto Fiorentino, Italy}

\author{G.~Roati}
\affiliation{Istituto Nazionale di Ottica del Consiglio Nazionale delle Ricerche (CNR-INO) c/o LENS, 50019 Sesto Fiorentino, Italy}
\affiliation{European Laboratory for Nonlinear Spectroscopy (LENS), University of Florence, 50019 Sesto Fiorentino, Italy}

\begin{abstract}

We 
investigate the role of vortices in the decay of persistent current states of annular atomic superfluids by solving numerically the Gross-Pitaevskii equation, and we directly compare our results with experimental data from  Ref.~\cite{PhysRevX.12.041037}. 
We theoretically model the optical phase-imprinting technique employed to experimentally excite finite-circulation states in Ref.~\cite{PhysRevX.12.041037} in the Bose-Einstein condensation regime, accounting for imperfections of the optical gradient imprinting profile. By comparing simulations of this realistic protocol to an ideal imprinting, we show that the introduced density excitations arising from imperfect imprinting are mainly responsible for limiting the maximum reachable winding number $w_\mathrm{max}$ in the superfluid ring. 
We also investigate the effect of a point-like obstacle with variable potential height $V_0$ onto the decay of circulating supercurrents. For a given obstacle height, a critical circulation $w_c$ exists, such that for an initial circulation $w_0$ larger than $w_c$ the supercurrent decays through the emission of 
vortices, which cross the superflow and thus induce phase slippage. Higher values of the obstacle height $V_0$ further favour the entrance of vortices, thus leading to lower values of $w_c$. Furthermore, the stronger vortex-defect interaction at higher $V_0$ leads to vortices that propagate closer to the center of the ring condensate. The combination of both these effects leads to an increase of the supercurrent decay rate for increasing $w_0$, in agreement with experimental observations \cite{PhysRevX.12.041037}. 

\end{abstract}

\maketitle


\section{Introduction}
One of the most fascinating properties of quantum fluids in multiply-connected geometries is the possibility to excite persistent currents, i.e. long-lived states of quantized circulation with a non-zero, dissipationless current \cite{Doll1961, Bluhm2009, Vinen1957, Ryu2007, Cai2021}.
In superconducting rings, persistent currents constitute the ground state of the system in the presence of an external magnetic field piercing the ring\cite{TinkhamBook}, whereas in neutral rotating superfluids they appear as metastable states in the energy spectrum \cite{Leggett1999, Mueller2002}.
A persistent current state with circulation $w$ corresponds to a wavefunction with an integer number $w$ of $2\pi$-phase windings. The quantized nature of the circulation makes such a state topologically protected against decaying into the underlying state with circulation $w-1$. In fact, the transition between the two states can occur only via a phase slip of $2\pi$ in the wavefunction. In rotating neutral superfluids, this results in an effective energy barrier separating circulation states with different $w$, creating the characteristic parabolic washboard energy spectrum \cite{Bloch1975}. 

Ultracold atomic superfluids in ring traps offer an ideal platform to investigate the stability of persistent currents, thanks to the high level of control of these system and to the possibility to introduce \textit{ad hoc} defects to trigger the current decay \cite{atomtronics}. 
A non-zero circulation can be introduced in an atomic superfluid ring in several ways: by stirring a repulsive obstacle through the ring with a well-defined frequency \cite{Wright2013_driving, Eckel2014_hysteresis, Cai2021}, by exciting Raman-transition with structured Laguerre-Gauss beam \cite{Beattie2013, Ramanathan2011}, or by optically phase-imprinting the desired winding onto the superfluid wavefunction \cite{Zheng2003, Perrin2018, PhysRevX.12.041037}. By accessing persistent current states at high $w$ and monitoring their long-time dynamics, the metastable character of persistent currents has been observed \cite{Moulder2012, Kumar2017}, and their decay has been induced by introducing a repulsive weak-link \cite{Wright2013_driving} or a localized obstacle in the ring \cite{PhysRevX.12.041037}. In these systems, the current decay proceeds via the emission of quantized vortices, that escape from the ring central hole, enter the bulk density and thus remove one circulation quantum from the local current.

In the recent work of Ref.~\cite{PhysRevX.12.041037}, we have investigated the stability of persistent currents of homogeneous paired fermionic superfluids in a clean ring trap and in the presence of a local defect.
We have accessed non-zero circulation states by phase imprinting an azimuthal gradient profile, which we have then probed interferometrically to measure the current in the ring. The phase-imprinting technique allowed us to effectively populate on-demand circulation state in all the different regimes throughout the crossover from a Bose-Einstein condensate (BEC) of tightly bound molecules to a Bardeen-Cooper-Shrieffer (BCS) superfluid of Cooper pairs. In particular, circulations up to $w=6$ have been obtained in the BEC limit with a single imprinting pulse, mainly limited by the unwanted excitations introduced by the non-perfectly sharp imprinting profile. 
On the other hand, in the presence of a single obstacle, we have observed the emergence of a critical circulation $w_c$: for $w<w_c$ the current is stable, while it decays via the emission of quantized vortices for $w>w_c$. While in the presence of the obstacle the connection between the current decay and the emission of vortices is apparent even in the experimental results, it is more elusive to investigate the source of the current instability in the clean ring, setting the upper limit for the highest $w$ which can be realized.

In this work, we numerically model the experiment reported in Ref.~\cite{PhysRevX.12.041037} in the BEC regime, to shed light on the microscopic mechanism connecting the persistent current instability to the emission of quantized vortices, both in the clean ring and in the presence of the obstacle. 
First, we investigate the effect of the non-ideal experimental imprinting procedure on the maximum winding number observed. By modelling the experimental imprinting profile, we investigate the stability of the excited current, confirming that the circulation is limited by the density excitations introduced by the imprinting, that favor the vortex emission. Then, we study the persistent current decay dynamics in the presence of a local obstacle in the ring density, as a function of the obstacle height. In all the studies reported in this paper, we observe the current decay to go along with the emission of quantized vortices and we characterized its microscopic dynamics. 


The paper is structured as follows. In Sect.~\ref{Sec:Methods} we describe the numerical methods employed for this study: in Sect.~\ref{Subec:imprinting} we present the different numerical techniques to excite the persistent currents, 
while in Sect.~\ref{Subec:methods_defect} we describe the numerical procedure  to study the decay induced by a local obstacle. 
In Sect.~\ref{Sec:Results} we report our results: in Sect.~\ref{Subec:observable} we compare the numerical and experimental technique to measure the persistent currents;
in Sect.~\ref{Subec:wmax} we discuss the maximum circulation state achievable under the experimental conditions; finally, in Sect.~\ref{Subec:obstacle} we report our numerical study of the defect-induced current decay as a function of its heights. 

\section{Methods}\label{Sec:Methods}
We study the dynamics of a Bose-Einstein condensate (BEC) in a ring geometry at $T=0$ by numerically solving the time-dependent mean-field 3D Gross-Pitaevskii Equation (GPE):

\begin{equation}
i \hbar \frac{\partial \Psi (\bold{r},t)}{\partial t} = - \frac{\hbar^2}{2 M} \nabla^2 \Psi  (\bold{r},t) +  V \Psi (\bold{r},t) + g | \Psi (\bold{r},t) |^2 \Psi (\bold{r},t)
\label{eq.GP.methods}
\end{equation}

where $\Psi(\bold{r},t)$ is the condensate wave function, $M$ the particle mass, $V$ the external trapping potential, and $g=4 \pi \hbar^2 a_M/M$ the interaction strength. 
To model the experiment reported in Ref.~\cite{PhysRevX.12.041037} in the BEC regime, we employ $M = 2\, m$, where $m$ is the mass of $^6$Li atoms, and $a_M = 0.6\, a$ is the molecular s-wave scattering length with $a = 1010 \, a_0$ the atomic one and $a_0$ the Bohr radius.
Throughout this work, the external trapping potential is provided by the combination of a 3D harmonic confinement and a hard-wall one in the $x-y$ plane, namely $V = V_{\rm harm} + V_{\rm ring}$, which creates a highly anisotropic superfluid, oblate in the vertical direction. The two are defined as follows:

\begin{equation}
V_{\rm harm} = \frac{1}{2} M \left ( \omega_\perp^2 r^2 + \omega_z^2 z^2 \right ),
\label{Vtrap}
\end{equation}

where $\{\omega_\perp , \, \omega_z \}= 2\pi \, \times \{2.5, \, 396\} \,\text{Hz}$ are the radial and axial trapping frequencies respectively, and

\begin{equation}
V_{\rm ring}=V_1 \left[\tanh\left(\frac{r-R_{\mathrm out}}{\sigma}\right)+1\right]+V_1 \left[\tanh\left(\frac{R_{\rm in}-r}{\sigma}\right)+1\right],
\label{Vring}
\end{equation}

where $\{R_{\rm in} , \, R_{\rm out} \}=  \{9.6 \, , \, 21.0\} \, \mu \text{m}$ are the inner and outer ring radii, $\sigma=0.37 \, \mu \text{m}$ and $V_1= 2.5 \, \mu$, where  $\mu$ is the chemical potential of the superfluid. The ring trap parameters $V_1$ and $\sigma$ have been chosen so to match the numerical density at equilibrium with the experimental one. 

To numerically find the system ground state, 
the equation~\ref{eq.GP.methods} is solved in imaginary time. In particular, we use a Cartesian grid composed of $\{ N_x , \, N_y , \, N_z \} = \{ 256 ,\, 256 , \, 80\}$ points along the $x$, $y$, and $z$ direction, respectively. We set the same grid size along the $x$- and $y$-axis, i.e.~$|x|, |y|<34.846 \, \mu $m and $|z|< 11.0 \, \mu$m.
The time step is  set to $\Delta t = 1 \times 10^{-5}\, \omega_\perp^{-1}$. 
For a particle number equal to the experimental one, namely $N=7.5\times 10^3$, we numerically obtain $\mu=1.06$ kHz and a healing length of $\xi = 0.61 \, \mu $m, consistent with the calculated ones for the experimental BEC superfluid \cite{PhysRevX.12.041037}. This gives rise to a grid spacing along the three directions of $\Delta x = \Delta y = 0.45 \, \xi$ and $\Delta z = 0.46 \, \xi$.

To directly compare with the experimental observables, we implement in our numerical simulations the interferometric technique employed experimentally to measure the circulation. It consists in introducing a disk condensate at the center of the superfluid ring as a phase reference and measuring the circulation in the ring from their interference pattern acquired in time-of-flight (TOF) \cite{Eckel_2014, Corman2014, Mathew2015}. To do so, we modify the trapping potential in the $x-y$ plane to account for the central disk condensate:
\begin{equation}
\displaystyle
V_{\rm ring}=V_1 \left[\tanh\left(\frac{r-R_{\rm out}}{\sigma} \right) +1 \right]+V_1 \exp \left[-\frac{2(R_{\rm in}-r)^2}{d ^2} \right].
\label{Vbound2}
\end{equation}
where  $d=1.1 \, \mu$m. The number of particles in such a modified trap is $ N = 10 \times 10^3$. 
To numerically obtain the interferograms, we remove $V_\mathrm{ring}$ in $0.3$ ms, while keeping the harmonic vertical confinement. As the superfluid expands in the $x-y$ plane during the TOF, we modify accordingly the numerical grid to contain it all while keeping the same grid spacing as for the simple ring simulations. 
Thus, we  modify our grid size by using $N_x=N_y=384$  grid points for a grid size of length  $|x|, |y|<53.186\, \mu$m. 


\subsection{Phase imprinting}\label{Subec:imprinting}

In the experiment of Ref.~\cite{PhysRevX.12.041037}, the current in the ring is excited via a phase imprinting technique: the superfluid density is illuminated for a time $t_I$ by a far-detuned laser with an intensity profile showing a linear gradient along the azimuthal direction. 
A phase $\Delta \phi (\theta) = U(\theta) \times t_I / \hbar$ is imprinted on the superfluid wavefunction, where $U(\theta)$ is the dipole potential exerted by the laser beam. The phase winding in the ring is thus directly controlled by acting on the imprinting time. Because of the finite resolution ($\sim 1\, \mu$m) of the optical setup imaging the imprinting pattern on the atomic cloud, the light gradient presents a sharp gradient in the opposite direction 
, which introduces density excitations.

Numerically, the winding number of the ring is imprinted by multiplying the initial wavefunction by the phase factor $\exp(-i\Delta \phi(\theta)) = \exp(-iU(\theta) t_I/ \hbar)$ during the imaginary time evolution.
In the following study, we have employed two different imprinting profile $U(\theta)$ to investigate different aspects of the persistent current stability. 
To examine the maximum winding number excitable under the experimental conditions, we choose $U(\theta)$ to closely model the experimental imprinting profile. 
In particular, for such a \textit{realistic} imprinting, we use:
\begin{equation}
U(\theta) = 
\begin{cases} 
      U_0 \left( 1- \frac{\theta}{2\pi - \Delta \theta}\right) & 0\leq \theta\leq 2\pi-\Delta \theta \\
      \frac{U_0}{\Delta \theta} \left( \theta- \Delta \theta +2\pi\right) & 2\pi -\Delta \theta < \theta < 2\pi,
   \end{cases}
\end{equation}
\label{Eq:real_imprinting}
where $U_0=7.8\, \mu$ is the height of the gradient at the discontinuity and $\Delta \theta=0.03$ rad. 

On the other hand, when studying the current stability in the presence of the obstacle, we employ an \textit{ideal} imprinting where $U(\theta)$ is a simple linear gradient, namely $U(\theta) = U_0 (1-\theta/2\pi)$. Such a procedure allows us to excite a well defined winding number $w_0$ without introducing any density excitations and therefore to decouple the current decay induced by the presence of the obstacle from the density excitation introduced by the imprinting. For both imprinting procedures, we define the total imprinted winding as $\Delta \phi_I = U_0 t_I/\hbar$, namely coinciding with the imprinted phase difference at the gradient discontinuity.

\subsection{Introducing a local obstacle}\label{Subec:methods_defect}


To study the dynamics of the ring currents in the presence of an obstacle, we solve Eq.~\ref{eq.GP.methods} by changing the trapping potential to account also for the defect. In particular, we add the defect contribution to the already presented expression of $V$, parametrizing it with the following Gaussian shape:
\begin{equation}
    \displaystyle
V_{\rm defect}=V_0  \exp \left[-\frac{2(x-x_0)^2}{\sigma_d^2} \right] \exp \left[-\frac{2(y-y_0)^2}{\sigma_d^2} \right]
\label{Eq:V_defect}
\end{equation}
where $(x_0, \,y_0)$ are the defect center coordinates along the $x$- and $y$-axis, $\sigma_d$ is its $1/e^2$ Gaussian width and $V_0$ its height. Throughout this work the defect is positioned almost central to the ring transverse extension, with $x_0=-15.0 \, \mu$m, and $y_0=0$. The defect width is kept fixed at $\sigma_d=1.4\, \mu$m while its height is varied in the range $0.8\, \mu \leq V_0 \leq 2.0 \, \mu$. 

In the numerical study of the obstacle-induced current decay, we modify the parameters of the ring trap, to account for the slightly different experimental conditions in the presence of the central disk used for the interferograms. In particular, we use $\{R_{\rm in} , \, R_{\rm out} \}=  \{10.45 , \, 21.0\} \, \mu \text{m}$, $\sigma=1.47 \, \mu \text{m}$ in Eq. \ref{Vring} and a total number of particles of $N=6035$. These parameters lead to a chemical potential $\mu=1.15\,$kHz. 
\section{Results}\label{Sec:Results}

\subsection{Numerical and experimental measurement of the ring winding number} \label{Subec:observable}

\begin{figure*}[ht!]
\centering
\vspace{0pt}
\includegraphics[width=0.65\textwidth]{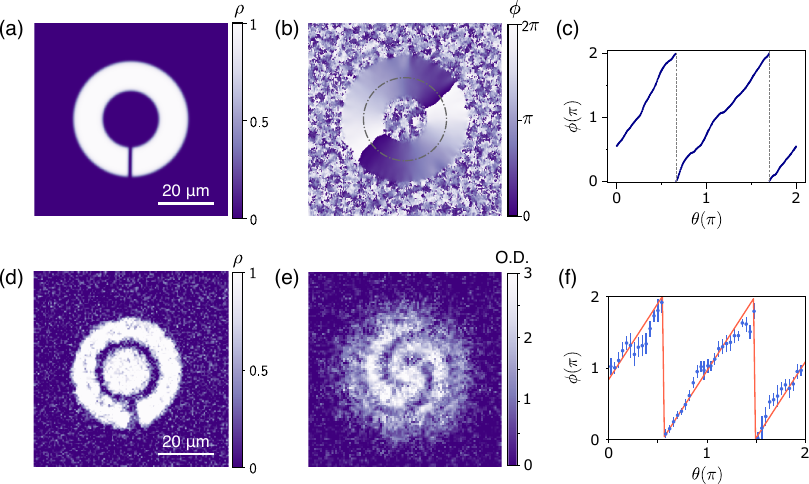}
\caption{Comparison between numerical and experimental observables. (\textbf{a}) Numerical density profile extracted immediately after a \textit{realistic} imprinting of $\Delta \phi_I = 2 \times 2\pi$. The 2D density is obtained by integrating $\rho(x,y,z)$ along the $z$-axis and scaling it to its maximum value.
(\textbf{b}) Numerical superfluid phase in the $xy$-plane $50\,$ms after the imprinting. 
(\textbf{c}) Azimuthal profile of the numerical phase profile as in (\textbf{b}) along the circle at $r= (R_{out}+R_{in})/2$ marked by the dash-dotted gray line.  
(\textbf{d}) Experimental \textit{in situ} density profile of the superfluid ring immediately after the imprinting of $\Delta \phi_I = 2\times 2\pi$.
(\textbf{e}) Experimental interferograms obtained $50\,$ms after the same imprinting as in (\textbf{d}).
(\textbf{f}) Azimuthal profile of the phase difference measured from the interferograms of (\textbf{e}). For each value of $\theta$, the local relative phase is here obtained from a sinusoidal fit of the radial density in the interferogram \cite{PhysRevX.12.041037}. The red solid line represent a linear fit, from the slope of which we extract the winding $w = 2.12(6)$.
\label{fig1}}
\end{figure*} 

From the solution of the GPE equation of Eq. \ref{eq.GP.methods} we extract the density $\rho$ of the atomic cloud and its phase $\phi$ by considering the Madelung representation of the wavefunction $\Psi(x,y,z)=\sqrt{\rho(x,y,z)} \exp(i\phi(x,y,z))$. In Fig.~\ref{fig1}(a, b), we report the 2D density and phase of the ring superfluid after a total imprinted winding of $\Delta \phi_I = 2\times 2\pi$ performed with the \textit{realistic} imprinting procedure. 
The superfluid ring shows a density depletion in correspondence of the gradient discontinuity, caused by the sharp anti-gradient in the imprinted profile of Eq.~\ref{Eq:real_imprinting}. The same perturbation is also observed in the experimental superfluid density reported in Fig.~\ref{fig1}d, acquired by imaging \textit{in situ} the atomic cloud immediately after the imprinting of $\Delta \phi_I = 2\times 2\pi$.
As it will be discussed in detail in the next section, the density excitations produced by the imperfections of the imprinting gradient 
set the effective limit on the maximum winding number we can excite.

While the 2D superfluid density provides informations on the imprinting-induced density excitations on the cloud, the phase profile of Fig.~\ref{fig1}b measures the circulation in the ring. The winding number of the superfluid ring is indeed encoded in the number of $2\pi$-jumps performed by the phase in the regions of non-zero density, where this quantity is well defined. In particular, Fig.~\ref{fig1}b shows $2$ phase jumps, as clearly illustrated in the azimuthal profile of $\phi$ reported in Fig.~\ref{fig1}c, signaling a winding number $w=2$ in the ring. More quantitatively, from the 2D profile of the superfluid phase we can calculate the mean circulation as $<w>= \int_{R_{in}}^{R_{out}} d\bold{r} \oint _{\bold{r}}\, \bold{v(\bold{r})} \, \cdot d\bold{l}$
, where $\textbf{v}(\textbf{r}) = \hbar/M \nabla \phi (\textbf{r})$ is the velocity of the superfluid. 

Experimentally, the phase of the ring is not an accessible quantity as for numerical simulations. However, the interferometric technique already discussed accesses the relative phase of the ring with respect to the disk superfluid at rest, such that from the interferograms we can measure the ring winding number. In Fig.~\ref{fig1}e we report the interferogram obtained after the imprinting of $\Delta \phi_I = 2\times 2\pi$. The $2$ clockwise spirals in such an image reveals a circulation of $w=2$ in the clockwise direction. From the sinusoidal fit of the radial profile of the interference pattern in the interferograms \cite{PhysRevX.12.041037}, the experimental azimuthal trend of the relative phase can be extracted (see Fig.~\ref{fig1}f). The local relative phase displays the same linear trend, modulus $2\pi$,
exhibited by the numerical phase of ~\ref{fig1}c, demonstrating that the two methods provides the same information on the winding number. We note that, as long as there are no vortices in the ring density, the mean circulation measured from the numerical phase profile coincides with the winding number we can extract form the interferogram, namely with the number of spirals appearing in the interference pattern.
However, when vortices are travelling in the ring density, the extracted value of the mean circulation is affected by the local phase winding around the vortices and the values of $\langle w \rangle$ extracted from the mean circulation and from the interferograms can slightly mismatch. 
In particular, the interferogram patterns measure the winding number of the internal edge of the ring, where the interference with the phase reference occurs, whereas the mean circulation accounts for the whole phase profile in the ring.

\subsection{The maximum circulation state}\label{Subec:wmax}

\begin{figure*}[t!]
\centering
\vspace{0pt}
\includegraphics[width=0.8\textwidth]{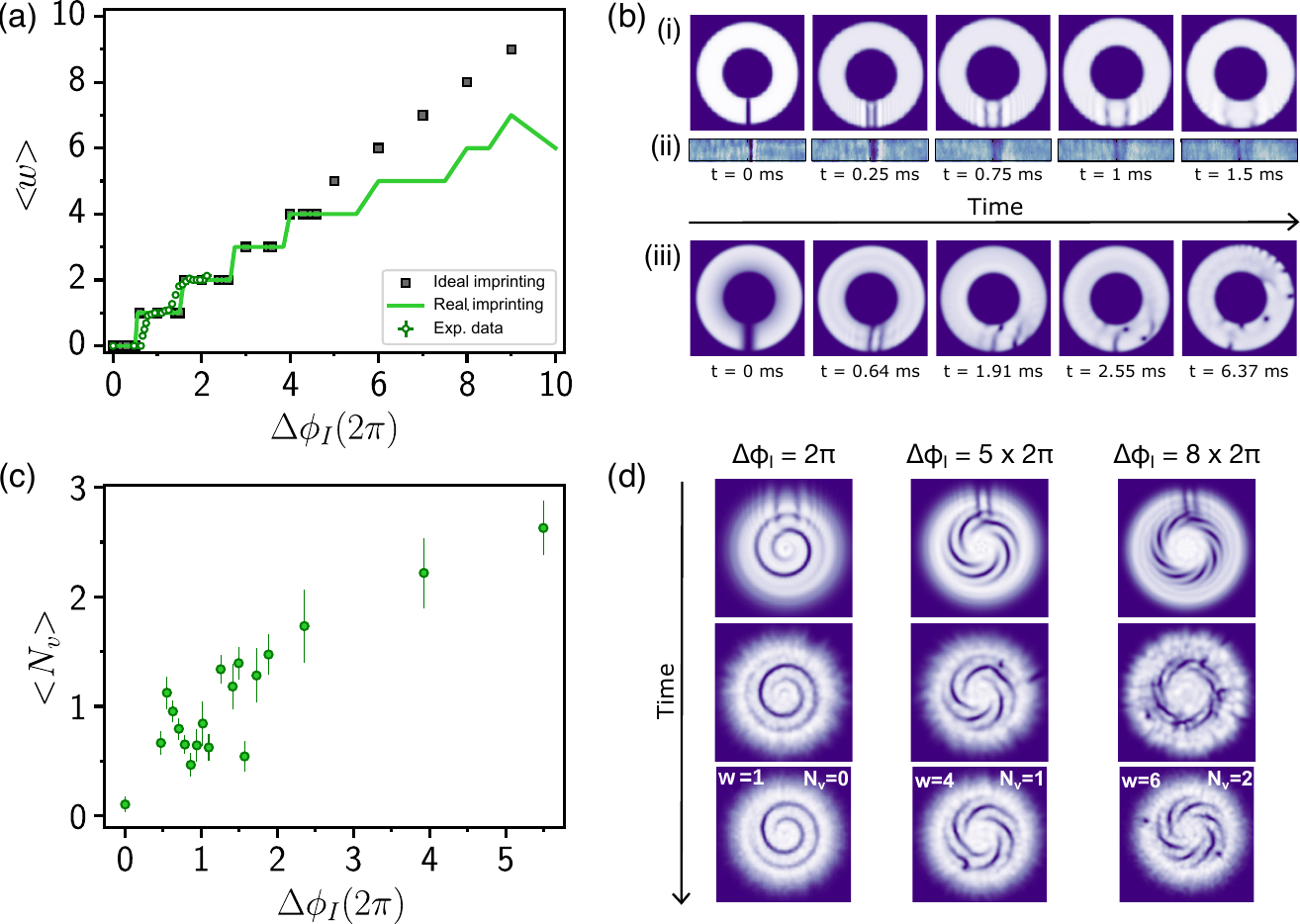}
\caption{(\textbf{a}) Average winding number $\langle w \rangle $ measured at time $t=50$ ms from the numerical (green line) and experimental (green symbols) interferograms  as a function of the imprinted phase $\Delta \phi _I$. The results of the numerical simulations with the \textit{ideal} imprinting are reported as black square, and they are performed only for integer values of $\Delta \phi_I / 2\pi$ for $\Delta \phi_I / 2\pi>5$. Experimental data are taken from Ref.~\cite{PhysRevX.12.041037}.
(\textbf{b}) Time evolution of the density excitation introduced by the imprinting for: (i) numerical data with \textit{realistic} imprinting of $\Delta \phi_I / 2 \pi= 1$, corresponding to $t_I = 127 \, \mu$s, (ii) experimental data with imprinting of $\Delta \phi_I / 2 \pi \simeq 1$ taken from Ref.~\cite{PhysRevX.12.041037}, (iii) numerical data with \textit{realistic} imprinting of $\Delta \phi_I / 2 \pi= 8$. Numerical density are plotted in cartesian coordines, the experimental one in polar coordinates.
(\textbf{c}) Number of vortices/antivortices $N_v$ experimentally observed in the ring density after $1.5\, $ms TOF \cite{PhysRevX.12.041037}. images are acquires $20\, $ms after the imprinting of the phase $\Delta \phi_I$. 
(\textbf{d}) Time evolution of the numerical inteferograms computed for imprinted phases of $\Delta \phi_I /2\pi = 1, \, 5, \, 8$. 
\label{fig2}}
\end{figure*}

The experimental data reported in Ref.~\cite{PhysRevX.12.041037} shows that the maximum circulation excitable with a single phase imprinting is $w_{ \rm max} = 6$ in the BEC regime. This value corresponds to superfluid velocities well below the speed of sound in every region of the ring, as the velocity of the superfluid at the inner ring radius is roughly $0.2 \, c_s$. To understand the origin of such a maximum winding number, we numerically study the imprinting procedure for both \textit{ideal} and \textit{realistic} imprinting. In both cases we measure the excited ring circulation from the numerical interferograms and compare it with the experimental imprinting characterization in Fig.~\ref{fig2}(a).
Both for experimental (green symbols) and numerical results obtained with any of the two imprinting methods (green line for \textit{realistic}, black symbols for \textit{ideal} imprinting), the measured winding $\langle w \rangle$ displays the typical step-like trend arising form the quantized nature of the circulation. However, whereas with the \textit{ideal} imprinting the measured winding keeps increasing, the excited circulation with the \textit{realistic} imprinting saturates at $w_{\rm max} = 7$ for large $\Delta \phi_I/ 2\pi \geq 7$, in qualitative agreement with the experimentally observed $w_{\rm max} = 6$ in the BEC regime \cite{PhysRevX.12.041037}. In particular, the numerical simulation for $\Delta \phi_I > 5$ with the \textit{realistic} imprinting show that initially a circulation of $\Delta \phi_I/2\pi$ is excited in the ring, but it quickly decays to a lower value in a timescale of a few 
ms, consistently with the experimental observations \cite{PhysRevX.12.041037}. Such a decay is never observed in the numerical results employing the \textit{ideal} imprinting, which is able to excite stable circulation state for arbitrarily large $\Delta \phi_I$. 
After an \textit{ideal} imprinting of a phase multiple  of $2\pi$, indeed the atomic density is observed to be unperturbed, with no density excitations that could limit the imprinted circulation.

On the other hand, both in the experimental \textit{in situ} density profile and in the numerical one after the \textit{realistic} imprinting procedure, a density depletion is clearly visible at the location of the sharp anti-gradient of the imprinting profile. 
In order to understand the connection of such a density perturbation with the observed $w_{\rm max}$,  
we follow its evolution in time. 
Figure~\ref{fig2}b shows the short-time dynamics of the ring superfluid density in the $xy$-plane for $\Delta \phi_I / 2\pi=1$ (i- numerical, ii- experimental) and for $\Delta \phi_I / 2\pi=8$ (iii). For small $\Delta \phi_I $, the initial density cut decays into two sound waves propagating in opposite directions, as observed both numerically and experimentally in Fig.~\ref{fig2}b-(i-ii). The presence of these sound waves only generates  some density fluctuations and therefore it does not affect the circulation in the ring, which is observed to be stable to $w=1$. 
On the other hand, for larger $\Delta \phi_I / 2\pi=8$, which corresponds to longer imprinting time, the extension of the initial density depletion increases (see Fig.~\ref{fig2}b-(iii), $t=0\,$ms). In this case, the decay of such a density perturbation leads to larger amplitude sound waves, which, favored also by the larger superfluid velocity, 
triggers the entrance of vortices both from the inner and from the outer edge of the ring. 
As a vortex enters the bulk density from the inner ring radius, it removes one circulation quantum from the local ring current, causing the imprinted circulation state to decay to a lower value. 
The presence of vortices in the superfluid density as a consequence of the imprinting is also observed experimentally, as reported in Fig.~\ref{fig2}c. In particular, we find that the average number of vortices $\langle N_v \rangle$ detected after the imprinting grows as a function of the imprinted phase difference and it is non-zero also for small $\Delta \phi_I$. 

The effect of the density-depletion decay on the circulation in the ring is clearly summarized in Fig.~\ref{fig2}d, where we report the numerical interferograms as a function of time after a \textit{realistic} imprinting of $\Delta \phi_I / 2\pi=1,\, 5, \, 8$. In the first case, the density depletion decay introduces only sound waves in the ring superfluid, which quickly decays out without affecting the circulation state, which is always measured to be $w=1$ in the interferograms. On the other hand, for larger $\Delta \phi_I$, the density depletion decay causes vortices to enter the ring superfluid density, which in turns induce the circulation state to decay to a lower value. In the case of $\Delta \phi_I / 2\pi=5$, the system reaches a new equilibrium in around $50$ ms, and the interferogram at this time shows the coexistence of a circulation of $w=4$ and one vortex. For $\Delta \phi_I / 2\pi=8$ instead, a larger time is needed to reach the new equilibrium, but eventually the system stabilizes to a circulation of $w=6$ with $2$ vortices, as shown in the interferogram at $t = 200 \,$ms. We note that in both $\Delta \phi_I / 2\pi=5, \, 8$ cases, the interferograms taken at intermediate time before reaching the equilibrium display a higher number of vortices, as they account also for the ones which entered the bulk from the outer ring, that do not affect the extracted ring winding number and quickly leave the ring density.

In conclusion, the density excitations introduced by the anti-gradient in the imprinting profile set the limit of the maximum circulation state we can populate.
Experimentally, it is observed that $w_\mathrm{max}$ increases with increasing $R_{in}$, confirming that the vortices emitted from the inner ring are responsible for the circulation state decay for large $\Delta \phi_I$. In fact, rings with smaller $R_{in}$ present a higher superfluid velocity at the inner ring radius, as $v \sim 1/r$. The critical velocity for vortex emission from the inner ring radius is therefore overcome for smaller $\Delta \phi_I$, leading to a lower maximum stable circulation state in the ring.

\begin{figure*}[t!]
\centering
\vspace{0pt}
\includegraphics[width=0.8\textwidth]{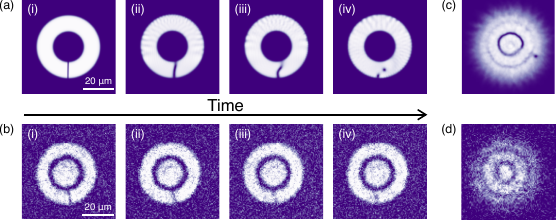}
\caption{(\textbf{a})-(\textbf{b}) Time evolution of the soliton-like density excitation introduced in the superfluid density after an imprinting of $\Delta \phi_I /2\pi= 0.49$ (a: numerical data with \textit{realistic} imprinting) and $\Delta \phi_I /2\pi= 0.54$ (b: experimental data). Numerical images are obtained after a time $0\,$ms (i), $3.4\,$ms (ii), $4.5\,$ms (iii), $6.2\,$ms (iv) from the imprinting; the experimental one are measured after $0\,$ms (i), $0.6\,$ms (ii), $1.4\,$ms (iii), $2.0\,$ms (iv) from the imprinting. The excitation decay dynamics happens over a faster timescale in the experiment, most likely because of its finite temperature.
(\textbf{c})-(\textbf{d}) Numerical and experimental interferograms acquired $50\,$ms after the same imprinting of panel (a) and (b), respectively. Both panels are adapted from Ref.~\cite{PhysRevX.12.041037}. 
\label{fig3}}
\end{figure*}   

Figure \ref{fig2}-(c) shows a non-zero number of vortices in the superfluid density even for $\Delta \phi_I /2 \pi<5$, when the imprinted circulation does not show any decay. These are mainly vortices entering the bulk from the outer ring, not perturbing the ring current, but they can also arise from the decay of a soliton-like excitation we introduce in the ring when imprinting $\Delta \phi_I = \pi +k \, 2\pi$, with $k$ integer. Under this condition, the depletion introduced by the imprinting has the phase profile of a soliton and similarly to a soliton it decays into a vortex-antivortex pair \cite{Donadello2014, ku2016cascade}. A comparison between the short-time dynamics for the numerical and experimental \textit{in situ} density profile after an imprinting of $\Delta \phi_I \simeq \pi$ is reported in Fig.~\ref{fig3}a,b.
In both cases we observe the soliton to bend, due to the velocity difference between the outer and inner edge of the superfluid, undergoing the so-called snake-instability to finally break into a vortex-antivortex pair \cite{anderson2001soliton, cetoli2013snake}, which have opposite circulation and propagate in opposite direction. 
Both the numerical and the experimental interferograms obtained $50$ ms after the imprinting measure $w=0$, signaled by an interference pattern with concentric rings [see Fig.~\ref{fig3}-(c-d)], but whereas in the numeric one the presence of a vortex is clearly evident, in the experimental one it is less striking.
In fact, because of its proximity to the outer edge of the ring, the antivortex is more sensible to the presence of thermal effects or other experimental fluctuations, that induce it to leave the bulk density faster in the experiment with respect to $T=0$ GPE simulations.
Anyway, a peak around $\langle N_v \rangle \simeq 1$ is clearly visible in Fig.~\ref{fig2}, as here vortices are counted $20\, $ms after the imprinting pulse, and are still present in the superfluid density.
We note that for an imprinting of $\Delta \phi_I = \pi + k \, 2\pi$, a soliton is excited also for the \textit{ideal} numerical imprinting procedure, showing a similar decay dynamics to the one illustrated in Fig.~\ref{fig2}. 
In conclusion, when imprinting a non-integer multiple of $2\pi$ the phase-structured density excitations produced at the gradient discontinuity always evolve into vortices, that then propagate on top of the ring current (if present), without further perturbing it.

\subsection{Supercurrent decay induced by the obstacle}\label{Subec:obstacle}

\begin{figure*}[t!]
\centering
\vspace{0pt}
\includegraphics[width=0.8\textwidth]{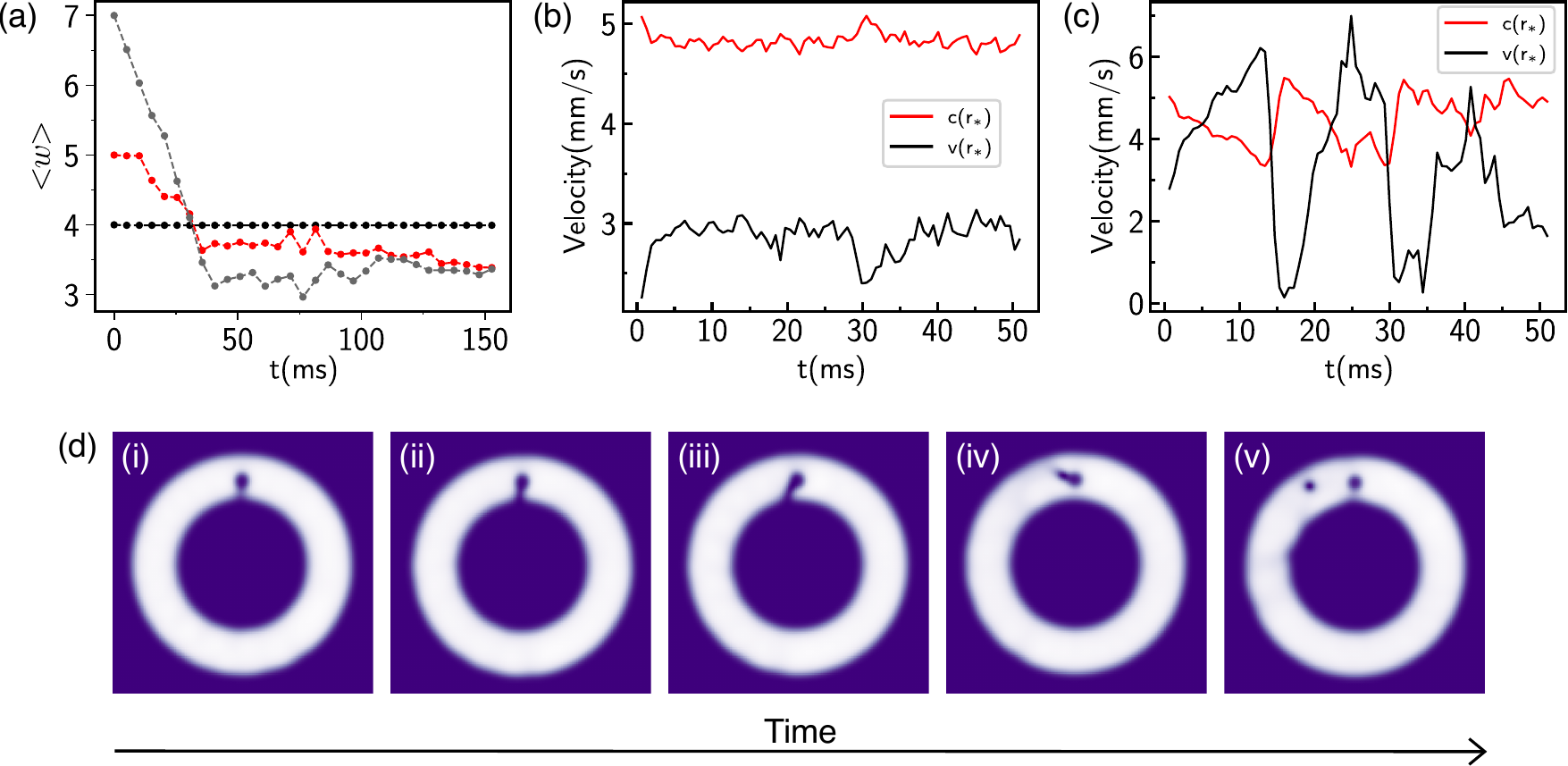}
\caption{(\textbf{a}) Time evolution of the mean circulation for $w_0=4, \, 5, \, 7$ in the presence of an obstacle. 
(\textbf{b}-\textbf{c}) Time evolution of the local superfluid velocity $v(r_*)$ and local speed of sound $c(r_*)$, extracted for $r_* = (R_{\rm in} +x_0)/2$, for $w_0=4=w_c$ (b) and $w_0=5>w_c$ (c).
(\textbf{d}) The 2D density profiles of the superfluid ring  extracted at different times following the vortex emission event for $w_0 = 5$; in particular the images are taken at $t=11.46$ ms (i), $t=14.01$ ms (ii), $t=15.28$  ms, $t=17.19$ ms and at $t=20.37$ ms. A low-density channel is created between the obstacle and the inner ring radius (i-iii), that favours the entrance of a vortex in the superfluid density (iv). After an interaction with the obstacle, the vortex leave the defect region and travels the superfluid density under the effect of the underlying current (iv-v). All the results reported in this figure are obtained for a defect height of $V_0/\mu=2$. 
\label{fig4}}
\end{figure*}

In this section we focus on the effect of an externally-introduced localized defect on the ring persistent current. Throughout this section, we employ the \textit{ideal} imprinting method in the numerical simulation in order to study the dissipation of the current introduced only by the obstacle.
We introduce the defect potential of Eq.~\ref{Eq:V_defect} at time $t=0$ and we monitor the evolution of the ring current, extracted from the 2D phase profile as the mean circulation. Figure~\ref{fig4}a shows the numerical results for an obstacle height of $V_0=2\,\mu$ and initial circulations of $w_0 = 4, \, 5, \, 7$.
Similarly to the experimental and theoretical findings of Ref.~\cite{PhysRevX.12.041037}, we observe the emergence of a critical circulation: initial states of $w_0\leq w_c$ are unperturbed by the obstacle, whereas for $w_0> w_c$ the current decays down to a lower circulation and it achieves its final value $w_f$ in timescales of a hundred ms. In particular, for $V_0=2\,\mu$ as in Fig.~\ref{fig4}a, the critical circulation is $w_c = 4$.
Consistently with the results reported in Ref.~\cite{PhysRevX.12.041037}, obtained for a smaller $V_0/\mu \simeq 1$, the supercurrent decay is observed to happen over faster timescale the larger is $w_0-w_c$.
Furthermore, as observed in Ref.~\cite{PhysRevX.12.041037} and as shown in Fig.~\ref{fig4}d, the current decay happens via the emission of quantized vortices, that enters the bulk superfluid through the low-density channel between the obstacle and the inner ring radius. 


To understand the vortex emission mechanism, we numerically compute the superfluid velocity $v$ and the local speed of sound $c$ at a point $r_*$ in the middle between the defect position and $R_{in}$. Figure \ref{fig4}-(b-c) show the results for the two values of $w_0=4=w_c$ (b) and $w_0=5>w_c$ (c). Consistently with the expected velocity trend $v(t=0) \approx w_0/r_*$, increasing the value of $w_0$ leads to larger values of the initial superfluid velocity. For $w_0 = 4$ we observe the local superfluid velocity, after a small initial increase, to saturate to a plateau value lower than $c$. However, for $w_0= 5 >w_c$, the creation of the low-density channel close to the defect leads the local superfluid velocity to increase until its value exceeds the local speed of sound $c (r_*)$, which on the other hand is decreasing, and a vortex enters the bulk density. When leaving the region at $r_*$, the vortex causes a phase-slip that decreases the superfluid velocity $v(r_*)$. Furthermore, in the region crossed by the vortex, the local circulation is decreased by one unit. 
In the $w_0 = 5$ case reported here, two vortices are emitted in correspondence of the first two superfluid velocity maxima, while the third one corresponds to the emission of only sound-waves. After this vortex emission dynamics, the superfluid  velocity reaches a smaller value than its initial one, i.e.~the ring current is dissipated.

\begin{figure*}[t!]
\centering
\vspace{0pt}
\includegraphics[width=0.65\textwidth]{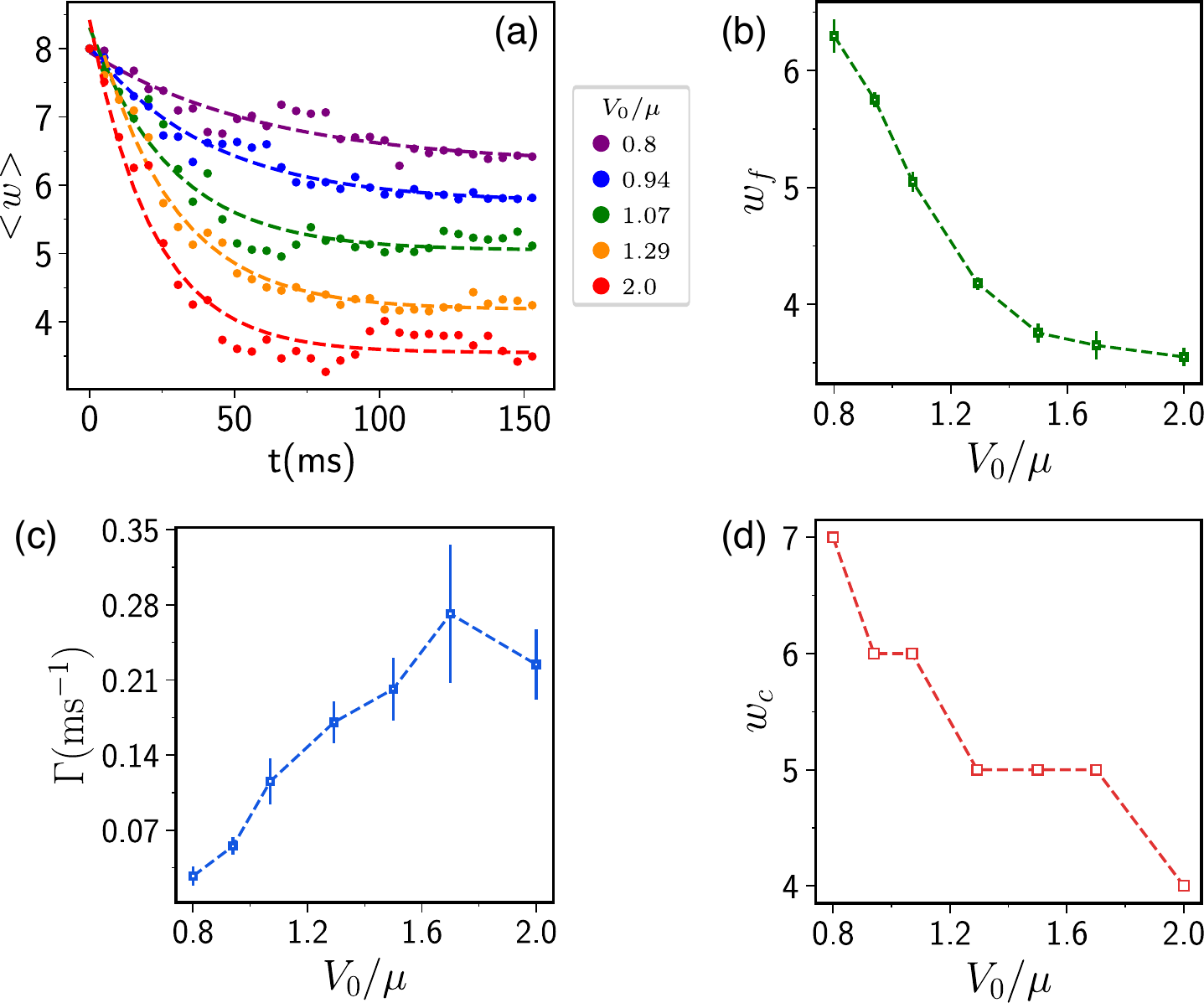}
\caption{(\textbf{a}) Time evolution of the mean circulation for $w_0=8$ in the presence of an obstacle of variable height $V_0/\mu$ (see legend). The dashed lines represent the fitted exponential profile for each defect height. (\textbf{b-c}) Final circulation $w_f$ (b) and winding number decay rate $\Gamma =a \times \gamma$ (c), as extracted from the exponential fit of the decays in panel a.
(\textbf{d}) Critical circulation $w_c$ as a function of the defect height $V_0/\mu$.   
\label{fig5}}
\end{figure*}

Finally, we report a numerical study of the effect of the vortex height on the current decay dynamics for $w_0=8$.
As reported in Fig.~\ref{fig5}a, increasing $V_0/\mu$ leads to a faster current decay to a lower final circulation. 
We quantitatively analyze the effect of the obstacle height by performing an exponential fit of the numerical data with the function  $f(t)=a \exp{(-\gamma t)}+w_f$ (dashed lines of Fig.~\ref{fig5}a). 
The fit results obtained for the final circulation $w_f$ are reported in Fig.~\ref{fig5}b. To compare the decay timescales under different $V_0/\mu$, we account for the different values of $w_f$ and define the winding decay rate $\Gamma = a\times \gamma$, plotted in Fig.~\ref{fig5}c. The  decrease of $w_f$ and the almost monotonic increase of $\Gamma$ in the explored range of $V_0/\mu$ results from the microscopic dynamics of the vortex emission process. 
In order to better understand their dependence on the defect height 
we extract the trend of the critical circulation $w_c$ as a function of the defect height, as reported in Fig.~\ref{fig5}d. We find that on average $w_c$ decreases with $V_0/\mu$, rapidly in the region of $V_0/\mu \simeq 1$ and more gently for higher obstacles. This leads to an increase of the vortex emission frequency for fixed $w_0$ as it increases with $w_0-w_c$ \cite{PhysRevX.12.041037}. In particular, for $V_0/\mu=0.8$ the vortices are emitted periodically roughly every $10\,$ms, while for $V_0/\mu=2$ the period for vortex emission is around $3\,$ms.  Furthermore, the total number of vortices emitted is in average larger for higher $V_0/\mu$, as higher $V_0$ determines a larger increase of the superfluid velocity in the channel between the obstacle and the inner ring radius.
More phase-slippage in the superfluid velocity are therefore needed to finally reduce the superflow below the critical value of the local speed of sound, leading to an increase decay rate because of the accumulative effect of  many vortices emission events \cite{Xhani2020}.
In addition, for large $V_0/\mu$, the stronger vortex-defect interaction induces each vortex to leave the obstacle region at a larger radius, thus yielding to a faster decrease of the mean circulation.

The effect of the defect in the current dynamics can be interpreted also under the parabolic washboard potential representation. That is, the obstacle in the superfluid density lowers the energy barrier $E_{b, w}$ separating the circulation states $w$ and $w-1$ from one another, such that $E_{b, w>w_c}=0$, which defines the critical circulation. For increasing defect height then, the barrier $E_{b, w}$ for a given $w$ gets lower, provoking the decrease of $w_c$. We can estimate the effect of the defect for different $V_0/\mu$ by calculating $E_{b, 1}$ from the numerical simulation results, as the state $w=1$ is stable under all the conditions we explored. In particular, we define $E_{b, 1} = E(\Delta \phi_I = \pi)-E(\Delta \phi_I = 2\pi)$, namely as the difference between the total energy of the system after the imprinting of $\pi$ and $2\pi$. We find that $E_{b, 1}(V_0/\mu=2)=0.62 \, E_{b, 1}(V_0/\mu=0.8)$, implying that increasing the defect height increases the probability  of a vortex to enter the ring superfluid. Interestingly, we note that the ratio of the energy barriers is compatible with the ratio between the observed $w_c$ at the two obstacle heights. 

\section{Conclusions}

We have performed a numerical study of the stability of persistent currents in a BEC under the experimental conditions of Ref.~\cite{PhysRevX.12.041037}. By numerically solving the GPE we have followed the dynamical evolution of currents, both in a clean ring and in the presence of a localized defect, identifying vortex emission as the source of the instability in both cases. 
In the clean ring, the density excitations introduced by the non-ideal imprinting favor vortices to enter the superflow from the ring center, effectively setting the upper limit of circulation which can be imprinted by a single optical imprinting pulse. 
In the presence of an obstacle, instead, 
for high enough $w$, the local superfluid velocity increases until its maximum value exceeds the local speed of sound, leading to vortex injection in the superfluid bulk that tears away a local circulation quantum. We found that a higher obstacle yield to a lower critical circulation and a faster decay. As a future perspective, it could be interesting to investigate how the presence of more obstacles, and their distribution over the superfluid density, would affect the stability of the current and its decay mechanism.

\begin{acknowledgements}
The authors acknowledge the LiLab experimental team at LENS, and especially Nicola Grani and Diego Hernandez-Rajkov, for providing data of the spurious vortices number, and Woo Jin Kwon, for stimulating discussions.
This work was supported by the European Research Council under Grant Agreement No. 307032, the Italian Ministry of University and Research under the PRIN2017 project CEnTraL,
and European Union’s Horizon 2020 research and innovation program under the Qombs project FET Flagship on Quantum Technologies Grant Agreement No. 820419. G.D.P. acknowledges financial support from the PNRR MUR project PE0000023-NQSTI.

\end{acknowledgements}

\bibliographystyle{apsrev4-1_ourstyle}

\begin{thebibliography}{25}%
\makeatletter
\providecommand \@ifxundefined [1]{%
 \@ifx{#1\undefined}
}%
\providecommand \@ifnum [1]{%
 \ifnum #1\expandafter \@firstoftwo
 \else \expandafter \@secondoftwo
 \fi
}%
\providecommand \@ifx [1]{%
 \ifx #1\expandafter \@firstoftwo
 \else \expandafter \@secondoftwo
 \fi
}%
\providecommand \natexlab [1]{#1}%
\providecommand \enquote  [1]{``#1''}%
\providecommand \bibnamefont  [1]{#1}%
\providecommand \bibfnamefont [1]{#1}%
\providecommand \citenamefont [1]{#1}%
\providecommand \href@noop [0]{\@secondoftwo}%
\providecommand \href [0]{\begingroup \@sanitize@url \@href}%
\providecommand \@href[1]{\@@startlink{#1}\@@href}%
\providecommand \@@href[1]{\endgroup#1\@@endlink}%
\providecommand \@sanitize@url [0]{\catcode `\\12\catcode `\$12\catcode
  `\&12\catcode `\#12\catcode `\^12\catcode `\_12\catcode `\%12\relax}%
\providecommand \@@startlink[1]{}%
\providecommand \@@endlink[0]{}%
\providecommand \url  [0]{\begingroup\@sanitize@url \@url }%
\providecommand \@url [1]{\endgroup\@href {#1}{\urlprefix }}%
\providecommand \urlprefix  [0]{URL }%
\providecommand \Eprint [0]{\href }%
\providecommand \doibase [0]{http://dx.doi.org/}%
\providecommand \selectlanguage [0]{\@gobble}%
\providecommand \bibinfo  [0]{\@secondoftwo}%
\providecommand \bibfield  [0]{\@secondoftwo}%
\providecommand \translation [1]{[#1]}%
\providecommand \BibitemOpen [0]{}%
\providecommand \bibitemStop [0]{}%
\providecommand \bibitemNoStop [0]{.\EOS\space}%
\providecommand \EOS [0]{\spacefactor3000\relax}%
\providecommand \BibitemShut  [1]{\csname bibitem#1\endcsname}%
\let\auto@bib@innerbib\@empty
\bibitem [{\citenamefont {Del~Pace}\ \emph {et~al.}(2022)\citenamefont
  {Del~Pace}, \citenamefont {Xhani}, \citenamefont {Muzi~Falconi},
  \citenamefont {Fedrizzi}, \citenamefont {Grani}, \citenamefont
  {Hernandez~Rajkov}, \citenamefont {Inguscio}, \citenamefont {Scazza},
  \citenamefont {Kwon},\ and\ \citenamefont {Roati}}]{PhysRevX.12.041037}%
  \BibitemOpen
  \bibfield  {author} {\bibinfo {author} {\bibfnamefont {G.}~\bibnamefont
  {Del~Pace}}, \bibinfo {author} {\bibfnamefont {K.}~\bibnamefont {Xhani}},
  \bibinfo {author} {\bibfnamefont {A.}~\bibnamefont {Muzi~Falconi}}, \bibinfo
  {author} {\bibfnamefont {M.}~\bibnamefont {Fedrizzi}}, \bibinfo {author}
  {\bibfnamefont {N.}~\bibnamefont {Grani}}, \bibinfo {author} {\bibfnamefont
  {D.}~\bibnamefont {Hernandez~Rajkov}}, \bibinfo {author} {\bibfnamefont
  {M.}~\bibnamefont {Inguscio}}, \bibinfo {author} {\bibfnamefont
  {F.}~\bibnamefont {Scazza}}, \bibinfo {author} {\bibfnamefont {W.~J.}\
  \bibnamefont {Kwon}}, \ and\ \bibinfo {author} {\bibfnamefont
  {G.}~\bibnamefont {Roati}},\ }\bibfield  {title} {\bibinfo {title} {\emph
  {Imprinting Persistent Currents in Tunable Fermionic Rings}},\ }\href
  {\doibase 10.1103/PhysRevX.12.041037} {\bibfield  {journal} {\bibinfo
  {journal} {Phys. Rev. X}\ }\textbf {\bibinfo {volume} {12}},\ \bibinfo
  {pages} {041037} (\bibinfo {year} {2022})}\BibitemShut {NoStop}%
\bibitem [{\citenamefont {Doll}\ and\ \citenamefont
  {N\"abauer}(1961)}]{Doll1961}%
  \BibitemOpen
  \bibfield  {author} {\bibinfo {author} {\bibfnamefont {R.}~\bibnamefont
  {Doll}}\ and\ \bibinfo {author} {\bibfnamefont {M.}~\bibnamefont
  {N\"abauer}},\ }\bibfield  {title} {\bibinfo {title} {\emph {Experimental
  Proof of Magnetic Flux Quantization in a Superconducting Ring}},\ }\href
  {\doibase 10.1103/PhysRevLett.7.51} {\bibfield  {journal} {\bibinfo
  {journal} {Phys. Rev. Lett.}\ }\textbf {\bibinfo {volume} {7}},\ \bibinfo
  {pages} {51} (\bibinfo {year} {1961})}\BibitemShut {NoStop}%
\bibitem [{\citenamefont {Bluhm}\ \emph {et~al.}(2009)\citenamefont {Bluhm},
  \citenamefont {Koshnick}, \citenamefont {Bert}, \citenamefont {Huber},\ and\
  \citenamefont {Moler}}]{Bluhm2009}%
  \BibitemOpen
  \bibfield  {author} {\bibinfo {author} {\bibfnamefont {H.}~\bibnamefont
  {Bluhm}}, \bibinfo {author} {\bibfnamefont {N.~C.}\ \bibnamefont {Koshnick}},
  \bibinfo {author} {\bibfnamefont {J.~A.}\ \bibnamefont {Bert}}, \bibinfo
  {author} {\bibfnamefont {M.~E.}\ \bibnamefont {Huber}}, \ and\ \bibinfo
  {author} {\bibfnamefont {K.~A.}\ \bibnamefont {Moler}},\ }\bibfield  {title}
  {\bibinfo {title} {\emph {Persistent Currents in Normal Metal Rings}},\
  }\href {\doibase 10.1103/PhysRevLett.102.136802} {\bibfield  {journal}
  {\bibinfo  {journal} {Phys. Rev. Lett.}\ }\textbf {\bibinfo {volume} {102}},\
  \bibinfo {pages} {136802} (\bibinfo {year} {2009})}\BibitemShut {NoStop}%
\bibitem [{\citenamefont {Vinen}(1957)}]{Vinen1957}%
  \BibitemOpen
  \bibfield  {author} {\bibinfo {author} {\bibfnamefont {W.~F.}\ \bibnamefont
  {Vinen}},\ }\bibfield  {title} {\bibinfo {title} {\emph {Mutual friction in a
  heat current in liquid helium II. II. Experiments on transient effects}},\
  }\href@noop {} {\bibfield  {journal} {\bibinfo  {journal} {Proceedings of the
  Royal Society of London. Series A. Mathematical and Physical Sciences}\
  }\textbf {\bibinfo {volume} {240}},\ \bibinfo {pages} {128} (\bibinfo {year}
  {1957})}\BibitemShut {NoStop}%
\bibitem [{\citenamefont {Ryu}\ \emph {et~al.}(2007)\citenamefont {Ryu},
  \citenamefont {Andersen}, \citenamefont {Clade}, \citenamefont {Natarajan},
  \citenamefont {Helmerson},\ and\ \citenamefont {Phillips}}]{Ryu2007}%
  \BibitemOpen
  \bibfield  {author} {\bibinfo {author} {\bibfnamefont {C.}~\bibnamefont
  {Ryu}}, \bibinfo {author} {\bibfnamefont {M.}~\bibnamefont {Andersen}},
  \bibinfo {author} {\bibfnamefont {P.}~\bibnamefont {Clade}}, \bibinfo
  {author} {\bibfnamefont {V.}~\bibnamefont {Natarajan}}, \bibinfo {author}
  {\bibfnamefont {K.}~\bibnamefont {Helmerson}}, \ and\ \bibinfo {author}
  {\bibfnamefont {W.~D.}\ \bibnamefont {Phillips}},\ }\bibfield  {title}
  {\bibinfo {title} {\emph {Observation of persistent flow of a Bose-Einstein
  condensate in a toroidal trap}},\ }\href@noop {} {\bibfield  {journal}
  {\bibinfo  {journal} {Phys. Rev. Lett.}\ }\textbf {\bibinfo {volume} {99}},\
  \bibinfo {pages} {260401} (\bibinfo {year} {2007})}\BibitemShut {NoStop}%
\bibitem [{\citenamefont {Cai}\ \emph {et~al.}(2022)\citenamefont {Cai},
  \citenamefont {Allman}, \citenamefont {Sabharwal},\ and\ \citenamefont
  {Wright}}]{Cai2021}%
  \BibitemOpen
  \bibfield  {author} {\bibinfo {author} {\bibfnamefont {Y.}~\bibnamefont
  {Cai}}, \bibinfo {author} {\bibfnamefont {D.~G.}\ \bibnamefont {Allman}},
  \bibinfo {author} {\bibfnamefont {P.}~\bibnamefont {Sabharwal}}, \ and\
  \bibinfo {author} {\bibfnamefont {K.~C.}\ \bibnamefont {Wright}},\ }\bibfield
   {title} {\bibinfo {title} {\emph {Persistent currents in rings of ultracold
  fermionic atoms}},\ }\href@noop {} {\bibfield  {journal} {\bibinfo  {journal}
  {Physical Review Letters}\ }\textbf {\bibinfo {volume} {128}},\ \bibinfo
  {pages} {150401} (\bibinfo {year} {2022})}\BibitemShut {NoStop}%
\bibitem [{\citenamefont {Tinkham}(2004)}]{TinkhamBook}%
  \BibitemOpen
  \bibfield  {author} {\bibinfo {author} {\bibfnamefont {M.}~\bibnamefont
  {Tinkham}},\ }\href@noop {} {\emph {\bibinfo {title} {Introduction to
  superconductivity}}}\ (\bibinfo  {publisher} {Courier Corporation},\ \bibinfo
  {year} {2004})\BibitemShut {NoStop}%
\bibitem [{\citenamefont {Leggett}(1999)}]{Leggett1999}%
  \BibitemOpen
  \bibfield  {author} {\bibinfo {author} {\bibfnamefont {A.~J.}\ \bibnamefont
  {Leggett}},\ }\bibfield  {title} {\bibinfo {title} {\emph {Superfluidity}},\
  }\href {\doibase 10.1103/RevModPhys.71.S318} {\bibfield  {journal} {\bibinfo
  {journal} {Rev. Mod. Phys.}\ }\textbf {\bibinfo {volume} {71}},\ \bibinfo
  {pages} {S318} (\bibinfo {year} {1999})}\BibitemShut {NoStop}%
\bibitem [{\citenamefont {Mueller}(2002)}]{Mueller2002}%
  \BibitemOpen
  \bibfield  {author} {\bibinfo {author} {\bibfnamefont {E.~J.}\ \bibnamefont
  {Mueller}},\ }\bibfield  {title} {\bibinfo {title} {\emph {Superfluidity and
  mean-field energy loops: Hysteretic behavior in Bose-Einstein condensates}},\
  }\href {\doibase 10.1103/PhysRevA.66.063603} {\bibfield  {journal} {\bibinfo
  {journal} {Phys. Rev. A}\ }\textbf {\bibinfo {volume} {66}},\ \bibinfo
  {pages} {063603} (\bibinfo {year} {2002})}\BibitemShut {NoStop}%
\bibitem [{\citenamefont {Bloch}(1973)}]{Bloch1975}%
  \BibitemOpen
  \bibfield  {author} {\bibinfo {author} {\bibfnamefont {F.}~\bibnamefont
  {Bloch}},\ }\bibfield  {title} {\bibinfo {title} {\emph {Superfluidity in a
  Ring}},\ }\href {\doibase 10.1103/PhysRevA.7.2187} {\bibfield  {journal}
  {\bibinfo  {journal} {Phys. Rev. A}\ }\textbf {\bibinfo {volume} {7}},\
  \bibinfo {pages} {2187} (\bibinfo {year} {1973})}\BibitemShut {NoStop}%
\bibitem [{\citenamefont {Amico}\ \emph {et~al.}(2021)\citenamefont {Amico},
  \citenamefont {Boshier}, \citenamefont {Birkl}, \citenamefont {Minguzzi},
  \citenamefont {Miniatura}, \citenamefont {Kwek}, \citenamefont {Aghamalyan},
  \citenamefont {Ahufinger}, \citenamefont {Anderson}, \citenamefont {Andrei},
  \citenamefont {Arnold}, \citenamefont {Baker}, \citenamefont {Bell},
  \citenamefont {Bland}, \citenamefont {Brantut}, \citenamefont {Cassettari},
  \citenamefont {Chetcuti}, \citenamefont {Chevy}, \citenamefont {Citro},
  \citenamefont {De~Palo}, \citenamefont {Dumke}, \citenamefont {Edwards},
  \citenamefont {Folman}, \citenamefont {Fortagh}, \citenamefont {Gardiner},
  \citenamefont {Garraway}, \citenamefont {Gauthier}, \citenamefont
  {G{\"u}nther}, \citenamefont {Haug}, \citenamefont {Hufnagel}, \citenamefont
  {Keil}, \citenamefont {Ireland}, \citenamefont {Lebrat}, \citenamefont {Li},
  \citenamefont {Longchambon}, \citenamefont {Mompart}, \citenamefont {Morsch},
  \citenamefont {Naldesi}, \citenamefont {Neely}, \citenamefont {Olshanii},
  \citenamefont {Orignac}, \citenamefont {Pandey}, \citenamefont
  {P{\'e}rez-Obiol}, \citenamefont {Perrin}, \citenamefont {Piroli},
  \citenamefont {Polo}, \citenamefont {Pritchard}, \citenamefont {Proukakis},
  \citenamefont {Rylands}, \citenamefont {Rubinsztein-Dunlop}, \citenamefont
  {Scazza}, \citenamefont {Stringari}, \citenamefont {Tosto}, \citenamefont
  {Trombettoni}, \citenamefont {Victorin}, \citenamefont {Klitzing},
  \citenamefont {Wilkowski}, \citenamefont {Xhani},\ and\ \citenamefont
  {Yakimenko}}]{atomtronics}%
  \BibitemOpen
  \bibfield  {author} {\bibinfo {author} {\bibfnamefont {L.}~\bibnamefont
  {Amico}}, \bibinfo {author} {\bibfnamefont {M.}~\bibnamefont {Boshier}},
  \bibinfo {author} {\bibfnamefont {G.}~\bibnamefont {Birkl}}, \bibinfo
  {author} {\bibfnamefont {A.}~\bibnamefont {Minguzzi}}, \bibinfo {author}
  {\bibfnamefont {C.}~\bibnamefont {Miniatura}}, \bibinfo {author}
  {\bibfnamefont {L.-C.}\ \bibnamefont {Kwek}}, \bibinfo {author}
  {\bibfnamefont {D.}~\bibnamefont {Aghamalyan}}, \bibinfo {author}
  {\bibfnamefont {V.}~\bibnamefont {Ahufinger}}, \bibinfo {author}
  {\bibfnamefont {D.}~\bibnamefont {Anderson}}, \bibinfo {author}
  {\bibfnamefont {N.}~\bibnamefont {Andrei}}, \bibinfo {author} {\bibfnamefont
  {A.~S.}\ \bibnamefont {Arnold}}, \bibinfo {author} {\bibfnamefont
  {M.}~\bibnamefont {Baker}}, \bibinfo {author} {\bibfnamefont {T.~A.}\
  \bibnamefont {Bell}}, \bibinfo {author} {\bibfnamefont {T.}~\bibnamefont
  {Bland}}, \bibinfo {author} {\bibfnamefont {J.~P.}\ \bibnamefont {Brantut}},
  \bibinfo {author} {\bibfnamefont {D.}~\bibnamefont {Cassettari}}, \bibinfo
  {author} {\bibfnamefont {W.~J.}\ \bibnamefont {Chetcuti}}, \bibinfo {author}
  {\bibfnamefont {F.}~\bibnamefont {Chevy}}, \bibinfo {author} {\bibfnamefont
  {R.}~\bibnamefont {Citro}}, \bibinfo {author} {\bibfnamefont
  {S.}~\bibnamefont {De~Palo}}, \bibinfo {author} {\bibfnamefont
  {R.}~\bibnamefont {Dumke}}, \bibinfo {author} {\bibfnamefont
  {M.}~\bibnamefont {Edwards}}, \bibinfo {author} {\bibfnamefont
  {R.}~\bibnamefont {Folman}}, \bibinfo {author} {\bibfnamefont
  {J.}~\bibnamefont {Fortagh}}, \bibinfo {author} {\bibfnamefont {S.~A.}\
  \bibnamefont {Gardiner}}, \bibinfo {author} {\bibfnamefont {B.~M.}\
  \bibnamefont {Garraway}}, \bibinfo {author} {\bibfnamefont {G.}~\bibnamefont
  {Gauthier}}, \bibinfo {author} {\bibfnamefont {A.}~\bibnamefont
  {G{\"u}nther}}, \bibinfo {author} {\bibfnamefont {T.}~\bibnamefont {Haug}},
  \bibinfo {author} {\bibfnamefont {C.}~\bibnamefont {Hufnagel}}, \bibinfo
  {author} {\bibfnamefont {M.}~\bibnamefont {Keil}}, \bibinfo {author}
  {\bibfnamefont {P.}~\bibnamefont {Ireland}}, \bibinfo {author} {\bibfnamefont
  {M.}~\bibnamefont {Lebrat}}, \bibinfo {author} {\bibfnamefont
  {W.}~\bibnamefont {Li}}, \bibinfo {author} {\bibfnamefont {L.}~\bibnamefont
  {Longchambon}}, \bibinfo {author} {\bibfnamefont {J.}~\bibnamefont
  {Mompart}}, \bibinfo {author} {\bibfnamefont {O.}~\bibnamefont {Morsch}},
  \bibinfo {author} {\bibfnamefont {P.}~\bibnamefont {Naldesi}}, \bibinfo
  {author} {\bibfnamefont {T.~W.}\ \bibnamefont {Neely}}, \bibinfo {author}
  {\bibfnamefont {M.}~\bibnamefont {Olshanii}}, \bibinfo {author}
  {\bibfnamefont {E.}~\bibnamefont {Orignac}}, \bibinfo {author} {\bibfnamefont
  {S.}~\bibnamefont {Pandey}}, \bibinfo {author} {\bibfnamefont
  {A.}~\bibnamefont {P{\'e}rez-Obiol}}, \bibinfo {author} {\bibfnamefont
  {H.}~\bibnamefont {Perrin}}, \bibinfo {author} {\bibfnamefont
  {L.}~\bibnamefont {Piroli}}, \bibinfo {author} {\bibfnamefont
  {J.}~\bibnamefont {Polo}}, \bibinfo {author} {\bibfnamefont {A.~L.}\
  \bibnamefont {Pritchard}}, \bibinfo {author} {\bibfnamefont {N.~P.}\
  \bibnamefont {Proukakis}}, \bibinfo {author} {\bibfnamefont {C.}~\bibnamefont
  {Rylands}}, \bibinfo {author} {\bibfnamefont {H.}~\bibnamefont
  {Rubinsztein-Dunlop}}, \bibinfo {author} {\bibfnamefont {F.}~\bibnamefont
  {Scazza}}, \bibinfo {author} {\bibfnamefont {S.}~\bibnamefont {Stringari}},
  \bibinfo {author} {\bibfnamefont {F.}~\bibnamefont {Tosto}}, \bibinfo
  {author} {\bibfnamefont {A.}~\bibnamefont {Trombettoni}}, \bibinfo {author}
  {\bibfnamefont {N.}~\bibnamefont {Victorin}}, \bibinfo {author}
  {\bibfnamefont {W.~v.}\ \bibnamefont {Klitzing}}, \bibinfo {author}
  {\bibfnamefont {D.}~\bibnamefont {Wilkowski}}, \bibinfo {author}
  {\bibfnamefont {K.}~\bibnamefont {Xhani}}, \ and\ \bibinfo {author}
  {\bibfnamefont {A.}~\bibnamefont {Yakimenko}},\ }\bibfield  {title} {\bibinfo
  {title} {\emph {{Roadmap on Atomtronics: State of the art and
  perspective}}},\ }\href {\doibase 10.1116/5.0026178} {\bibfield  {journal}
  {\bibinfo  {journal} {AVS Quantum Sci.}\ }\textbf {\bibinfo {volume} {3}}
  (\bibinfo {year} {2021}),\ 10.1116/5.0026178},\ \bibinfo {note}
  {039201}\BibitemShut {NoStop}%
\bibitem [{\citenamefont {Wright}\ \emph {et~al.}(2013)\citenamefont {Wright},
  \citenamefont {Blakestad}, \citenamefont {Lobb}, \citenamefont {Phillips},\
  and\ \citenamefont {Campbell}}]{Wright2013_driving}%
  \BibitemOpen
  \bibfield  {author} {\bibinfo {author} {\bibfnamefont {K.~C.}\ \bibnamefont
  {Wright}}, \bibinfo {author} {\bibfnamefont {R.}~\bibnamefont {Blakestad}},
  \bibinfo {author} {\bibfnamefont {C.~J.}\ \bibnamefont {Lobb}}, \bibinfo
  {author} {\bibfnamefont {W.~D.}\ \bibnamefont {Phillips}}, \ and\ \bibinfo
  {author} {\bibfnamefont {G.~K.}\ \bibnamefont {Campbell}},\ }\bibfield
  {title} {\bibinfo {title} {\emph {Driving phase slips in a superfluid atom
  circuit with a rotating weak link}},\ }\href@noop {} {\bibfield  {journal}
  {\bibinfo  {journal} {Phys. Rev. Lett.}\ }\textbf {\bibinfo {volume} {110}},\
  \bibinfo {pages} {025302} (\bibinfo {year} {2013})}\BibitemShut {NoStop}%
\bibitem [{\citenamefont {Eckel}\ \emph
  {et~al.}(2014{\natexlab{a}})\citenamefont {Eckel}, \citenamefont {Lee},
  \citenamefont {Jendrzejewski}, \citenamefont {Murray}, \citenamefont {Clark},
  \citenamefont {Lobb}, \citenamefont {Phillips}, \citenamefont {Edwards},\
  and\ \citenamefont {Campbell}}]{Eckel2014_hysteresis}%
  \BibitemOpen
  \bibfield  {author} {\bibinfo {author} {\bibfnamefont {S.}~\bibnamefont
  {Eckel}}, \bibinfo {author} {\bibfnamefont {J.~G.}\ \bibnamefont {Lee}},
  \bibinfo {author} {\bibfnamefont {F.}~\bibnamefont {Jendrzejewski}}, \bibinfo
  {author} {\bibfnamefont {N.}~\bibnamefont {Murray}}, \bibinfo {author}
  {\bibfnamefont {C.~W.}\ \bibnamefont {Clark}}, \bibinfo {author}
  {\bibfnamefont {C.~J.}\ \bibnamefont {Lobb}}, \bibinfo {author}
  {\bibfnamefont {W.~D.}\ \bibnamefont {Phillips}}, \bibinfo {author}
  {\bibfnamefont {M.}~\bibnamefont {Edwards}}, \ and\ \bibinfo {author}
  {\bibfnamefont {G.~K.}\ \bibnamefont {Campbell}},\ }\bibfield  {title}
  {\bibinfo {title} {\emph {Hysteresis in a quantized superfluid
  ‘atomtronic’circuit}},\ }\href@noop {} {\bibfield  {journal} {\bibinfo
  {journal} {Nature}\ }\textbf {\bibinfo {volume} {506}},\ \bibinfo {pages}
  {200} (\bibinfo {year} {2014}{\natexlab{a}})}\BibitemShut {NoStop}%
\bibitem [{\citenamefont {Beattie}\ \emph {et~al.}(2013)\citenamefont
  {Beattie}, \citenamefont {Moulder}, \citenamefont {Fletcher},\ and\
  \citenamefont {Hadzibabic}}]{Beattie2013}%
  \BibitemOpen
  \bibfield  {author} {\bibinfo {author} {\bibfnamefont {S.}~\bibnamefont
  {Beattie}}, \bibinfo {author} {\bibfnamefont {S.}~\bibnamefont {Moulder}},
  \bibinfo {author} {\bibfnamefont {R.~J.}\ \bibnamefont {Fletcher}}, \ and\
  \bibinfo {author} {\bibfnamefont {Z.}~\bibnamefont {Hadzibabic}},\ }\bibfield
   {title} {\bibinfo {title} {\emph {Persistent currents in spinor
  condensates}},\ }\href@noop {} {\bibfield  {journal} {\bibinfo  {journal}
  {Phys. Rev. Lett.}\ }\textbf {\bibinfo {volume} {110}},\ \bibinfo {pages}
  {025301} (\bibinfo {year} {2013})}\BibitemShut {NoStop}%
\bibitem [{\citenamefont {Ramanathan}\ \emph {et~al.}(2011)\citenamefont
  {Ramanathan}, \citenamefont {Wright}, \citenamefont {Muniz}, \citenamefont
  {Zelan}, \citenamefont {Hill~III}, \citenamefont {Lobb}, \citenamefont
  {Helmerson}, \citenamefont {Phillips},\ and\ \citenamefont
  {Campbell}}]{Ramanathan2011}%
  \BibitemOpen
  \bibfield  {author} {\bibinfo {author} {\bibfnamefont {A.}~\bibnamefont
  {Ramanathan}}, \bibinfo {author} {\bibfnamefont {K.}~\bibnamefont {Wright}},
  \bibinfo {author} {\bibfnamefont {S.~R.}\ \bibnamefont {Muniz}}, \bibinfo
  {author} {\bibfnamefont {M.}~\bibnamefont {Zelan}}, \bibinfo {author}
  {\bibfnamefont {W.}~\bibnamefont {Hill~III}}, \bibinfo {author}
  {\bibfnamefont {C.}~\bibnamefont {Lobb}}, \bibinfo {author} {\bibfnamefont
  {K.}~\bibnamefont {Helmerson}}, \bibinfo {author} {\bibfnamefont
  {W.}~\bibnamefont {Phillips}}, \ and\ \bibinfo {author} {\bibfnamefont
  {G.}~\bibnamefont {Campbell}},\ }\bibfield  {title} {\bibinfo {title} {\emph
  {Superflow in a toroidal Bose-Einstein condensate: an atom circuit with a
  tunable weak link}},\ }\href@noop {} {\bibfield  {journal} {\bibinfo
  {journal} {Phys. Rev. Lett.}\ }\textbf {\bibinfo {volume} {106}},\ \bibinfo
  {pages} {130401} (\bibinfo {year} {2011})}\BibitemShut {NoStop}%
\bibitem [{\citenamefont {Zheng}\ and\ \citenamefont
  {Javanainen}(2003)}]{Zheng2003}%
  \BibitemOpen
  \bibfield  {author} {\bibinfo {author} {\bibfnamefont {Y.}~\bibnamefont
  {Zheng}}\ and\ \bibinfo {author} {\bibfnamefont {J.}~\bibnamefont
  {Javanainen}},\ }\bibfield  {title} {\bibinfo {title} {\emph {Classical and
  quantum models for phase imprinting}},\ }\href@noop {} {\bibfield  {journal}
  {\bibinfo  {journal} {Phys. Rev. A}\ }\textbf {\bibinfo {volume} {67}},\
  \bibinfo {pages} {035602} (\bibinfo {year} {2003})}\BibitemShut {NoStop}%
\bibitem [{\citenamefont {Kumar}\ \emph {et~al.}(2018)\citenamefont {Kumar},
  \citenamefont {Dubessy}, \citenamefont {Badr}, \citenamefont {De~Rossi},
  \citenamefont {de~Go\"er~de Herve}, \citenamefont {Longchambon},\ and\
  \citenamefont {Perrin}}]{Perrin2018}%
  \BibitemOpen
  \bibfield  {author} {\bibinfo {author} {\bibfnamefont {A.}~\bibnamefont
  {Kumar}}, \bibinfo {author} {\bibfnamefont {R.}~\bibnamefont {Dubessy}},
  \bibinfo {author} {\bibfnamefont {T.}~\bibnamefont {Badr}}, \bibinfo {author}
  {\bibfnamefont {C.}~\bibnamefont {De~Rossi}}, \bibinfo {author}
  {\bibfnamefont {M.}~\bibnamefont {de~Go\"er~de Herve}}, \bibinfo {author}
  {\bibfnamefont {L.}~\bibnamefont {Longchambon}}, \ and\ \bibinfo {author}
  {\bibfnamefont {H.}~\bibnamefont {Perrin}},\ }\bibfield  {title} {\bibinfo
  {title} {\emph {Producing superfluid circulation states using phase
  imprinting}},\ }\href {\doibase 10.1103/PhysRevA.97.043615} {\bibfield
  {journal} {\bibinfo  {journal} {Phys. Rev. A}\ }\textbf {\bibinfo {volume}
  {97}},\ \bibinfo {pages} {043615} (\bibinfo {year} {2018})}\BibitemShut
  {NoStop}%
\bibitem [{\citenamefont {Moulder}\ \emph {et~al.}(2012)\citenamefont
  {Moulder}, \citenamefont {Beattie}, \citenamefont {Smith}, \citenamefont
  {Tammuz},\ and\ \citenamefont {Hadzibabic}}]{Moulder2012}%
  \BibitemOpen
  \bibfield  {author} {\bibinfo {author} {\bibfnamefont {S.}~\bibnamefont
  {Moulder}}, \bibinfo {author} {\bibfnamefont {S.}~\bibnamefont {Beattie}},
  \bibinfo {author} {\bibfnamefont {R.~P.}\ \bibnamefont {Smith}}, \bibinfo
  {author} {\bibfnamefont {N.}~\bibnamefont {Tammuz}}, \ and\ \bibinfo {author}
  {\bibfnamefont {Z.}~\bibnamefont {Hadzibabic}},\ }\bibfield  {title}
  {\bibinfo {title} {\emph {Quantized supercurrent decay in an annular
  Bose-Einstein condensate}},\ }\href@noop {} {\bibfield  {journal} {\bibinfo
  {journal} {Phys. Rev. A}\ }\textbf {\bibinfo {volume} {86}},\ \bibinfo
  {pages} {013629} (\bibinfo {year} {2012})}\BibitemShut {NoStop}%
\bibitem [{\citenamefont {Kumar}\ \emph {et~al.}(2017)\citenamefont {Kumar},
  \citenamefont {Eckel}, \citenamefont {Jendrzejewski},\ and\ \citenamefont
  {Campbell}}]{Kumar2017}%
  \BibitemOpen
  \bibfield  {author} {\bibinfo {author} {\bibfnamefont {A.}~\bibnamefont
  {Kumar}}, \bibinfo {author} {\bibfnamefont {S.}~\bibnamefont {Eckel}},
  \bibinfo {author} {\bibfnamefont {F.}~\bibnamefont {Jendrzejewski}}, \ and\
  \bibinfo {author} {\bibfnamefont {G.~K.}\ \bibnamefont {Campbell}},\
  }\bibfield  {title} {\bibinfo {title} {\emph {Temperature-induced decay of
  persistent currents in a superfluid ultracold gas}},\ }\href@noop {}
  {\bibfield  {journal} {\bibinfo  {journal} {Phys. Rev. A}\ }\textbf {\bibinfo
  {volume} {95}},\ \bibinfo {pages} {021602} (\bibinfo {year}
  {2017})}\BibitemShut {NoStop}%
\bibitem [{\citenamefont {Eckel}\ \emph
  {et~al.}(2014{\natexlab{b}})\citenamefont {Eckel}, \citenamefont
  {Jendrzejewski}, \citenamefont {Kumar}, \citenamefont {Lobb},\ and\
  \citenamefont {Campbell}}]{Eckel_2014}%
  \BibitemOpen
  \bibfield  {author} {\bibinfo {author} {\bibfnamefont {S.}~\bibnamefont
  {Eckel}}, \bibinfo {author} {\bibfnamefont {F.}~\bibnamefont
  {Jendrzejewski}}, \bibinfo {author} {\bibfnamefont {A.}~\bibnamefont
  {Kumar}}, \bibinfo {author} {\bibfnamefont {C.~J.}\ \bibnamefont {Lobb}}, \
  and\ \bibinfo {author} {\bibfnamefont {G.~K.}\ \bibnamefont {Campbell}},\
  }\bibfield  {title} {\bibinfo {title} {\emph {Interferometric measurement of
  the current-phase relationship of a superfluid weak link}},\ }\href@noop {}
  {\bibfield  {journal} {\bibinfo  {journal} {Phys. Rev. X}\ }\textbf {\bibinfo
  {volume} {4}},\ \bibinfo {pages} {031052} (\bibinfo {year}
  {2014}{\natexlab{b}})}\BibitemShut {NoStop}%
\bibitem [{\citenamefont {Corman}\ \emph {et~al.}(2014)\citenamefont {Corman},
  \citenamefont {Chomaz}, \citenamefont {Bienaim{\'e}}, \citenamefont
  {Desbuquois}, \citenamefont {Weitenberg}, \citenamefont {Nascimbene},
  \citenamefont {Dalibard},\ and\ \citenamefont {Beugnon}}]{Corman2014}%
  \BibitemOpen
  \bibfield  {author} {\bibinfo {author} {\bibfnamefont {L.}~\bibnamefont
  {Corman}}, \bibinfo {author} {\bibfnamefont {L.}~\bibnamefont {Chomaz}},
  \bibinfo {author} {\bibfnamefont {T.}~\bibnamefont {Bienaim{\'e}}}, \bibinfo
  {author} {\bibfnamefont {R.}~\bibnamefont {Desbuquois}}, \bibinfo {author}
  {\bibfnamefont {C.}~\bibnamefont {Weitenberg}}, \bibinfo {author}
  {\bibfnamefont {S.}~\bibnamefont {Nascimbene}}, \bibinfo {author}
  {\bibfnamefont {J.}~\bibnamefont {Dalibard}}, \ and\ \bibinfo {author}
  {\bibfnamefont {J.}~\bibnamefont {Beugnon}},\ }\bibfield  {title} {\bibinfo
  {title} {\emph {Quench-induced supercurrents in an annular Bose gas}},\
  }\href@noop {} {\bibfield  {journal} {\bibinfo  {journal} {Phys. Rev. Lett.}\
  }\textbf {\bibinfo {volume} {113}},\ \bibinfo {pages} {135302} (\bibinfo
  {year} {2014})}\BibitemShut {NoStop}%
\bibitem [{\citenamefont {Mathew}\ \emph {et~al.}(2015)\citenamefont {Mathew},
  \citenamefont {Kumar}, \citenamefont {Eckel}, \citenamefont {Jendrzejewski},
  \citenamefont {Campbell}, \citenamefont {Edwards},\ and\ \citenamefont
  {Tiesinga}}]{Mathew2015}%
  \BibitemOpen
  \bibfield  {author} {\bibinfo {author} {\bibfnamefont {R.}~\bibnamefont
  {Mathew}}, \bibinfo {author} {\bibfnamefont {A.}~\bibnamefont {Kumar}},
  \bibinfo {author} {\bibfnamefont {S.}~\bibnamefont {Eckel}}, \bibinfo
  {author} {\bibfnamefont {F.}~\bibnamefont {Jendrzejewski}}, \bibinfo {author}
  {\bibfnamefont {G.~K.}\ \bibnamefont {Campbell}}, \bibinfo {author}
  {\bibfnamefont {M.}~\bibnamefont {Edwards}}, \ and\ \bibinfo {author}
  {\bibfnamefont {E.}~\bibnamefont {Tiesinga}},\ }\bibfield  {title} {\bibinfo
  {title} {\emph {Self-heterodyne detection of the in situ phase of an atomic
  superconducting quantum interference device}},\ }\href@noop {} {\bibfield
  {journal} {\bibinfo  {journal} {Phys. Rev. A}\ }\textbf {\bibinfo {volume}
  {92}},\ \bibinfo {pages} {033602} (\bibinfo {year} {2015})}\BibitemShut
  {NoStop}%
\bibitem [{\citenamefont {Donadello}\ \emph {et~al.}(2014)\citenamefont
  {Donadello}, \citenamefont {Serafini}, \citenamefont {Tylutki}, \citenamefont
  {Pitaevskii}, \citenamefont {Dalfovo}, \citenamefont {Lamporesi},\ and\
  \citenamefont {Ferrari}}]{Donadello2014}%
  \BibitemOpen
  \bibfield  {author} {\bibinfo {author} {\bibfnamefont {S.}~\bibnamefont
  {Donadello}}, \bibinfo {author} {\bibfnamefont {S.}~\bibnamefont {Serafini}},
  \bibinfo {author} {\bibfnamefont {M.}~\bibnamefont {Tylutki}}, \bibinfo
  {author} {\bibfnamefont {L.~P.}\ \bibnamefont {Pitaevskii}}, \bibinfo
  {author} {\bibfnamefont {F.}~\bibnamefont {Dalfovo}}, \bibinfo {author}
  {\bibfnamefont {G.}~\bibnamefont {Lamporesi}}, \ and\ \bibinfo {author}
  {\bibfnamefont {G.}~\bibnamefont {Ferrari}},\ }\bibfield  {title} {\bibinfo
  {title} {\emph {Observation of solitonic vortices in Bose-Einstein
  condensates}},\ }\href@noop {} {\bibfield  {journal} {\bibinfo  {journal}
  {Physical review letters}\ }\textbf {\bibinfo {volume} {113}},\ \bibinfo
  {pages} {065302} (\bibinfo {year} {2014})}\BibitemShut {NoStop}%
\bibitem [{\citenamefont {Ku}\ \emph {et~al.}(2016)\citenamefont {Ku},
  \citenamefont {Mukherjee}, \citenamefont {Yefsah},\ and\ \citenamefont
  {Zwierlein}}]{ku2016cascade}%
  \BibitemOpen
  \bibfield  {author} {\bibinfo {author} {\bibfnamefont {M.~J.}\ \bibnamefont
  {Ku}}, \bibinfo {author} {\bibfnamefont {B.}~\bibnamefont {Mukherjee}},
  \bibinfo {author} {\bibfnamefont {T.}~\bibnamefont {Yefsah}}, \ and\ \bibinfo
  {author} {\bibfnamefont {M.~W.}\ \bibnamefont {Zwierlein}},\ }\bibfield
  {title} {\bibinfo {title} {\emph {Cascade of solitonic excitations in a
  superfluid fermi gas: From planar solitons to vortex rings and lines}},\
  }\href@noop {} {\bibfield  {journal} {\bibinfo  {journal} {Physical review
  letters}\ }\textbf {\bibinfo {volume} {116}},\ \bibinfo {pages} {045304}
  (\bibinfo {year} {2016})}\BibitemShut {NoStop}%
\bibitem [{\citenamefont {Xhani}\ \emph {et~al.}(2020)\citenamefont {Xhani},
  \citenamefont {Neri}, \citenamefont {Galantucci}, \citenamefont {Scazza},
  \citenamefont {Burchianti}, \citenamefont {Lee}, \citenamefont {Barenghi},
  \citenamefont {Trombettoni}, \citenamefont {Inguscio}, \citenamefont
  {Zaccanti}, \citenamefont {Roati},\ and\ \citenamefont
  {Proukakis}}]{Xhani2020}%
  \BibitemOpen
  \bibfield  {author} {\bibinfo {author} {\bibfnamefont {K.}~\bibnamefont
  {Xhani}}, \bibinfo {author} {\bibfnamefont {E.}~\bibnamefont {Neri}},
  \bibinfo {author} {\bibfnamefont {L.}~\bibnamefont {Galantucci}}, \bibinfo
  {author} {\bibfnamefont {F.}~\bibnamefont {Scazza}}, \bibinfo {author}
  {\bibfnamefont {A.}~\bibnamefont {Burchianti}}, \bibinfo {author}
  {\bibfnamefont {K.-L.}\ \bibnamefont {Lee}}, \bibinfo {author} {\bibfnamefont
  {C.~F.}\ \bibnamefont {Barenghi}}, \bibinfo {author} {\bibfnamefont
  {A.}~\bibnamefont {Trombettoni}}, \bibinfo {author} {\bibfnamefont
  {M.}~\bibnamefont {Inguscio}}, \bibinfo {author} {\bibfnamefont
  {M.}~\bibnamefont {Zaccanti}}, \bibinfo {author} {\bibfnamefont
  {G.}~\bibnamefont {Roati}}, \ and\ \bibinfo {author} {\bibfnamefont {N.~P.}\
  \bibnamefont {Proukakis}},\ }\bibfield  {title} {\bibinfo {title} {\emph
  {Critical Transport and Vortex Dynamics in a Thin Atomic Josephson
  Junction}},\ }\href {\doibase 10.1103/PhysRevLett.124.045301} {\bibfield
  {journal} {\bibinfo  {journal} {Phys. Rev. Lett.}\ }\textbf {\bibinfo
  {volume} {124}},\ \bibinfo {pages} {045301} (\bibinfo {year}
  {2020})}\BibitemShut {NoStop}%
\end{thebibliography}

\end{document}